\documentclass[preprint,aps]{revtex4}
\usepackage{graphicx}
\usepackage{dcolumn}
\begin{document}

\title{New Physics effects on  decay $B_s \to \gamma\gamma$ in Technicolor Model }

\author{%
Qin XiuMei\email{snowlotusqin@yahoo.com.cn} }
\author{%
Wujun Huo\email{whuo@ictp.it}}
\author{%
Xiaofang Yang}

\affiliation{
Department of Physics Department, Southeast University, Nanjing,
Jiangsu 211189, China}

\begin{abstract}
In this paper we calculate the contributions to the branching ratio
of $B_s \to \gamma\gamma$ from the charged Pseudo-Goldstone bosons
appeared in one generation Technicolor model. We find that the
theoretical values of the branching ratio, $BR(B_s\to\gamma\gamma)$,
including the contributions of PGBs, $P^\pm$ and $P^{\pm}_8$  , are
much different from the $SM$ prediction. The new physics effects can
be enhance 2-3 levels to $SM$ result. It is shown that the decay
$B_s\to \gamma\gamma$ can give the test the new physics signals from
the technicolor model.
\end{abstract}


\maketitle

\section{introduction}

As is well known, the rare radiative decays of $B$ mesons is in
particular sensitive to contributions from those new physics beyond
the standard model(SM). Both inclusive and exclusive processes, such
as the decays $B_s \to X\gamma$, $B_s \to \gamma\gamma$ and $B \to
X_s\gamma$ have been received some attention in the
literature$^{[1-14]}$. In this paper, we will present our results in
Technicolor theories.

The one generation Technicolor model (OGTM)$^{[15-16]}$is the
simplest and most frequently studied model which contained the
parameters are less than SM. Same as other models, the OGTM has its
defects such as the S parameter large and positive$^{[17]}$. But we
can relax the constraints on the OGTM form the $S$ parameter by
introducing three additional parameters $(V,W,X)$$^{[18]}$. The
basic idea of the OGTM is: we introduce a new set of asymptotically
free gauge interactions and the Technicolor force act on
Technifermions. The Technicolor interaction at $1Tev$ become strong
and cause a spontaneous breaking of the global flavor symmetry
$SU(8)_L \times SU(8)_R\times U(1)_{Y}$. The result is $8^{2}-1=63$
massless Goldstone bosons. Three of the these objects replace the
Higgs field and induce a mass of $W^{\pm}$ and $Z^0$ gauge bosons.
And at the new strong interaction other Goldstone bosons acquire
masses. As for the $B_s \to \gamma\gamma$, only the charged color
single and color octets have contributions. The gauge couplings of
the PGBs are determined by their quantum numbers. In Table 1 we
listed the relevant couplings$^{[19]}$ needed in our calculation,
where the $V_{ud} $ is the corresponding element of
$Kobayashi-Maskawa$ matrix . The Goldstone boson decay constant
$F_\pi$$^{[20]}$ should be $F_{\pi}=v/2=123GeV$, which corresponds
to the vacuum expectation of an elementary Higgs field .

\begin{table}[htb]
\begin{center}
\begin{tabular}{|c|c|c|c|c|c|c|c|c|}
\hline
$P^+ P^- \gamma_\mu$  & $-ie(p_+ - p_-)_\mu$ \\
\hline $P^+_{8a} P^-_{8b} \gamma_\mu$  & $-ie(p_+ - p_-)_\mu
\delta_{ab}$ \\\hline
$P^+\; u\; d$  & $i\frac{V_{ud}}{2
F_\pi}\sqrt{\frac{2}{3}} [M_u (1-\gamma_5) - M_d (1 + \gamma_5) ]$
\\ \hline
$P^+_{8a}\; u\; d$  & $i\frac{V_{ud}}{2 F_\pi}  \lambda_a [M_u
(1-\gamma_5) - M_d (1 + \gamma_5) ]$ \\ \hline $P^+_{8a} P^-_{8b}
g_{c\mu}$  & $-g f_{abc}(p_a - p_b)_\mu $ \\
\hline 
\end{tabular}
\end{center}
\label{sm4}
\caption{The relevant gauge couplings and Effective
Yukawa couplings for the OGTM.}
\end{table}

At the LO in QCD the effective Hamiltonian is
\begin{equation}
{\cal H}_{eff} =\frac{-4G_F}{\sqrt{2}} V_{tb}V_{ts}^*
        \displaystyle{\sum_{i=1}^{8} }C_i (M_W^-) O_i(M_W^-).
\end{equation}
Where, as usual, $G_{F}$ denotes the Fermi coupling constant and
$V_{tb}V_{ts}^*$ indicates the Cabibbo-Kobayashi-Maskawa matrix
element.And the current-current, QCD penguin, electromagnetic and
chromomagnetic dipole operators are of the form
\begin{eqnarray}
O_1&=&(\overline{c}_{L\beta} \gamma^{\mu} b_{L\alpha})
            (\overline{s}_{L\alpha} \gamma_{\mu} c_{L\beta})\;\\
O_2&=&(\overline{c}_{L\alpha} \gamma^{\mu} b_{L\alpha})
            (\overline{s}_{L\beta} \gamma_{\mu} c_{L\beta})\;\\
O_3&=&(\overline{s}_{L\alpha} \gamma^{\mu} b_{L\alpha})
\sum_{q=u,d,s,c,b}(\overline{q}_{L\beta} \gamma_{\mu} q_{L\beta})\;\\
O_4&=&(\overline{s}_{L\alpha} \gamma^{\mu} b_{L\beta})
\sum_{q=u,d,s,c,b}(\overline{q}_{L\beta} \gamma_{\mu} q_{L\alpha})\;\\
O_5&=&(\overline{s}_{L\alpha} \gamma^{\mu} b_{L\alpha})
\sum_{q=u,d,s,c,b}(\overline{q}_{R\beta} \gamma_{\mu} q_{R\beta})\;\\
O_6&=&(\overline{s}_{L\alpha} \gamma^{\mu} b_{L\beta})
\sum_{q=u,d,s,c,b}(\overline{q}_{R\beta} \gamma_{\mu} q_{R\alpha})\;\\
O_7&=&(e/16\pi^2) m_b \overline{s}_L \sigma^{\mu\nu}
            b_{R} F_{\mu\nu}\;\\
O_8&=&(g/16\pi^2) m_b \overline{s}_{L} \sigma^{\mu\nu}
            T^a b_{R} G_{\mu\nu}^a\;
\end{eqnarray}
where $\alpha$ and $\beta$ are color indices, $\alpha=1, . . . ,8$
labels SU(3)c generators, e and $g$ refer to the electromagnetic and
strong coupling constants, while $F_{\mu\nu}$ and $G^{a}_{\mu\nu}$
denote the QED and QCD field strength tensors, respectively.

The Feynman diagrams that contribute to the matrix element as the
following
\begin{figure}[th]
{\includegraphics{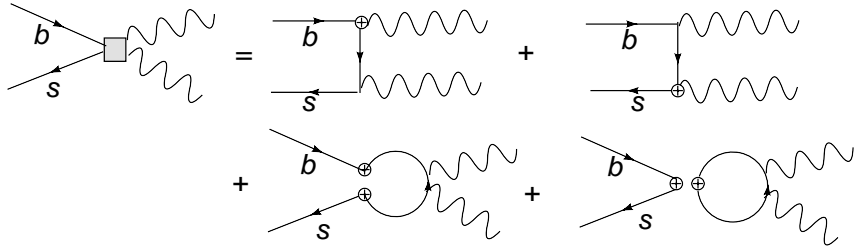}}
\caption
{Examples of Feynman diagrams that contribute to the matrix element.} \label{G1}
\end{figure}
\begin{figure}[th]
{\includegraphics{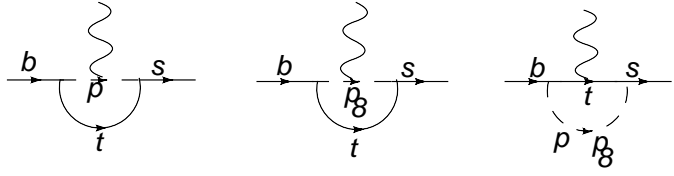}}
\caption
{The Feynman diagrams that contribute to the
Wilson coefficients C7,C8.} \label{G2}
\end{figure}
In Fig.2 the shot-dash lines represent the charged PGBs $P^\pm$ and
$P^{\pm}_8$ of OGTM. We at first integrate out the top quark and the
weak $W$ bosons at $\mu=M_{W}$ scale, generating an effective
five-quark theory and run the effective field theory down to b-quark
scale to give the leading log QCD corrections by using the
renormalization group equation. The Wilson coefficients are process
independent and the coefficients $C_{i}(\mu)$ of 8 operators are
calculated from the Fig.2.The Wilson coefficients are read$^{[21]}$
\begin{eqnarray}
C_i(M_W)=0, \;\; i=1,3,4,5,6, \;\;\; C_2(M_W)=1,\\
C_7(M_W)=-A(\delta) +\frac{B(x)}{3\sqrt{2}G_F F_{\pi}^2 } +\frac{8
B(y)}{3\sqrt{2}G_F F_{\pi}^2 } \label{c7}\\
C_8(M_W)=-C(\delta)+\frac{D(x)}{3\sqrt{2}G_F F_{\pi}^2}+\frac{8D(y)
+ E(y)}{3\sqrt{2}G_F F_{\pi}^2 }\label{c8}
\end{eqnarray}

with $\delta=M_W^2/m_t^2$, $x=(m(P^{\pm})/m_t)^2$ and
$y=(m(P^{\pm}_8)/m_t)^2$.From the $Eq(11),(12)$ , we can see the
situation of the color-octet charged PGBs is more complicate than
that of the color-singlet charged PGBs ,because of the involvement
of the color interactions. where
\begin{eqnarray}
A(\delta)&=&\frac{ \frac{1}{3} +\frac{5}{24} \delta -\frac{7}{24}
        \delta^2}{(1-\delta)^3}
        +\frac{ \frac{3}{4}\delta -\frac{1}{2}\delta^2}{(1-\delta)^4}
\log[\delta] \\
B(y)& =& \frac{ -\frac{11}{36}
        +\frac{53}{72}y -\frac{25}{72}y^2}{(1-y)^3}\nonumber \\
        &+&\frac{ -\frac{1}{4}y +\frac{2}{3}y^2 -\frac{1}{3}y^3}
        {(1-y)^4}\log[y]\\
C(\delta)&=&\frac{\frac{1}{8} -\frac{5}{8} \delta-\frac{1}{4}
        \delta^2}{(1-\delta)^3}
        -\frac{ \frac{3}{4}\delta^2}{(1-\delta)^4} \log[\delta] \\
D(y)& =&\frac{ -\frac{5}{24}
        +\frac{19}{24}y -\frac{5}{6}y^2}{(1-y)^3}\nonumber \\
        &+&\frac{ \frac{1}{4}y^2 -\frac{1}{2}y^3}{(1-y)^4}
        \log[y]\\
E(y) & =&\frac{ \frac{3}{2}-\frac{15}{8}y -\frac{15}{8}y^2
}{(1-y)^3}
        +\frac{\frac{9}{4}y -\frac{9}{2}y^2}{(1-y)^4 }\log[y]
\end{eqnarray}
By caculate the graphs of the exchanged $W$ boson in the SM we
gained the function $A$ and $C$;And by caculate the graphs of the
exchanged color-singlet and color-octet charged PGBs in OGTM we
gained the function $B$, $D$ and $E$. when $\delta < 1$, $x,y >> 1$,
the OGTM contribution $B$, $D$ and $E$ have always a relative minus
sign with the SM contribution $A$ and $C$. As a result, the OGTM
contribution always destructively interferes with the SM
contribution.

The leading-order results for the Wilson coefficients of all
operators entering the effective Hamiltonian in Eq.(1) can be
written in an analytic form. They are
\begin{eqnarray}
C_7^{eff}(m_b) &=& \eta^{16/23}C_7(M_W) +\frac{8}{3} (
\eta^{14/23}-\eta^{16/23} )\times \nonumber \\&&C_8(M_W)+C_2(M_W)
\displaystyle \sum _{i=1}^{8} h_i \eta^{a_i}.
\end{eqnarray}
With $\eta = \alpha_s(M_W) /\alpha_s (m_b)$,
\begin{eqnarray}
h_i &=&(\frac{626126}{272277}, -\frac{56281}{51730}, -\frac{3}{7},
-\frac{1}{14},-0.6494,\nonumber \\&& -0.0380, -0.0186, -0.0057
).\\
a_i &=&(\frac{14}{23}, \frac{16}{23}, \frac{6}{23},
-\frac{12}{23},\nonumber \\&& 0.4086, -0.4230, -0.8994, 0.1456 ).
\end{eqnarray}

To calculate $B_s \to \gamma\gamma$ , one may follow a perturbative
QCD approach which includes a proof of factorization, showing that
soft gluon effects can be factorized into $B_{s}$ meson wave
function; and a systematic way of resumming large logarithms due to
hard gluons with energies between 1Gev and $m_{b}$. In order to
calculate the matrix element of Eq(1) for the $B_s \to \gamma\gamma$
, we can work in the weak binding approximation and assume that both
the $b$ and the $s$ quarks are at rest in the $B_s $ meson, and the
$b$ quarks carries most of the meson energy, and its four velocity
can be treated as equal to that of $B_s $. Hence one may write $b$
quark momentum as $p_{b}=m_{b}v$ where is the common four velocity
of $b$ and $B_{s}$. We have
\begin{eqnarray}
p_{b}\cdot k_1&=&m_bv\cdot k_1={1\over 2}m_bm_{B_s}=p_{b}\cdot k_2,\nonumber \\
p_{s}\cdot k_1&=&(p-k_1-k_2)\cdot k_1=\nonumber \\&&-{1\over2}
m_{B_s}(m_{B_s}-m_b)=p_{s}\cdot k_2,
\end{eqnarray}

We compute the amplitude of $B_s \to \gamma\gamma$ using the
following relations
\begin{eqnarray}
\left\langle 0\vert \bar{s}\gamma_{\mu}\gamma_5 b\vert B_s(P)
\right\rangle
&=& -if_{B_s}P_{\mu},\nonumber \\
\left\langle 0\vert \bar{s}\gamma_5 b\vert B_s(P) \right\rangle &=&
if_{B_s}M_B,
\end{eqnarray}
where $f_{B_s}$ is the $B_s$ meson decay constant which is about
$200$ MeV .

The total amplitude is now separated into a CP-even and a CP-odd
part
\begin{equation}
T(B_s\to \gamma\gamma)=M^+F_{\mu\nu}F^{\mu\nu}
+iM^-F_{\mu\nu}\tilde{F}^{\mu\nu}.
\end{equation}
We find that
\begin{eqnarray}
M^+&=&{-4{\sqrt 2}\alpha G_F\over
9\pi}f_{B_s}m_{b_{s}}V_{ts}^*V_{tb}\times\nonumber \\&&
\left(\frac{m_b}{m_{B_s}}B K(m_b^2) +{3C_7\over 8\bar{\Lambda}
}\right).
\end{eqnarray}
with $B= -(3C_6+C_5)/4$, $ \bar{\Lambda}=m_{B_s}-m_b$, and
\begin{eqnarray}
M^-&=&{4{\sqrt 2}\alpha G_F\over
9\pi}f_{B_s}m_{b_{s}}V_{ts}^*V_{tb}\times\nonumber \\&& \left(\sum_q
A_qJ(m_q^2)+ \frac{m_b}{m_{B_s}}BL(m_b^2)+{3C_7\over 8\bar{\Lambda}}
\right).
\end{eqnarray}
where
\begin{eqnarray}
A_u &=&(C_3-C_5)N_c+(C_4-C_6)\nonumber \\
A_d &=&{1\over 4}\left[(C_3-C_5)N_c+(C_4-C_6)\right]\nonumber \\
A_c &=&(C_1+C_3-C_5)N_c+(C_2+C_4-C_6) \nonumber \\
A_s &=&{1\over 4}\left[(C_3+C_4-C_5)N_c+(C_3+C_4-C_6)\right]\\
A_s &=&{1\over 4}\left[(C_3+C_4-C_5)N_c+(C_3+C_4-C_6)\right].
\end{eqnarray}
The functions $J(m^2)$, $K(m^2)$ and $L(m^2)$  are defined by
\begin{eqnarray}
J(m^2)&=&I_{11}(m^2),\nonumber \\
K(m^2)&=&4(I_{11}(m^2)-I_{00}(m^2)) ,\nonumber \\
L(m^2)&=&I_{00}(m^2),
\end{eqnarray}
with
\begin{equation}
I_{pq}(m^2)=\int_{0}^{1}{dx}\int_{0}^{1-x}{dy}\frac{x^{p}y^{q}}{m^{2}-2xyk_{1}\cdot
k_{2}-i\varepsilon}
\end{equation}

The decay width for $B_s\to \gamma\gamma$ is simply
\begin{equation}
\Gamma(B_s\to \gamma\gamma)={m_{B_s}^3\over 16\pi}({\vert M^+\vert
}^2+{\vert M^-\vert }^2).
\end{equation}

In SM, with $C_{2}=C_{2}(M_{W})=1$ , and the other Wilson
coefficients are zero, we find $\Gamma( B_s\to
\gamma\gamma)=1.3\times 10^{-10} \ {\rm eV}$ which amounts to a
branching ratio $Br(B_s\to \gamma\gamma)=3.5\times 10^{-7}$, for the
given $\Gamma^{total}_{B_s}=4\times 10^{-4} \  {\rm eV}$. In
numerical calculations we use the corresponding input parameters
$M_W=80.22\;GeV$, $\alpha_s(m_Z)=0.117$, $m_c=1.5\;GeV$,
$m_b=4.8\;GeV$ and $|V_{tb} V_{ts}^*|^2/ |V_{cb}|^2= 0.95$ ,
respectively. The present experimental limit$^{[22]}$ on the decay
$B_s\to\gamma\gamma$ is
\begin{eqnarray}
{\rm Br}(B_s \to\gamma\gamma)\leq 8.6\times 10^{-6},
\end{eqnarray} which is far from the theoretical results. So, we can not put the
constraint to the masses of PGBs. The constraints of the masses of
$P^\pm$ and $P^{\pm}_8$ can be from the decay$^{[24]}$ $B\to
s\gamma$ : $m_{P^\pm_8} >400$GeV.

\begin{figure}[th]
{\includegraphics{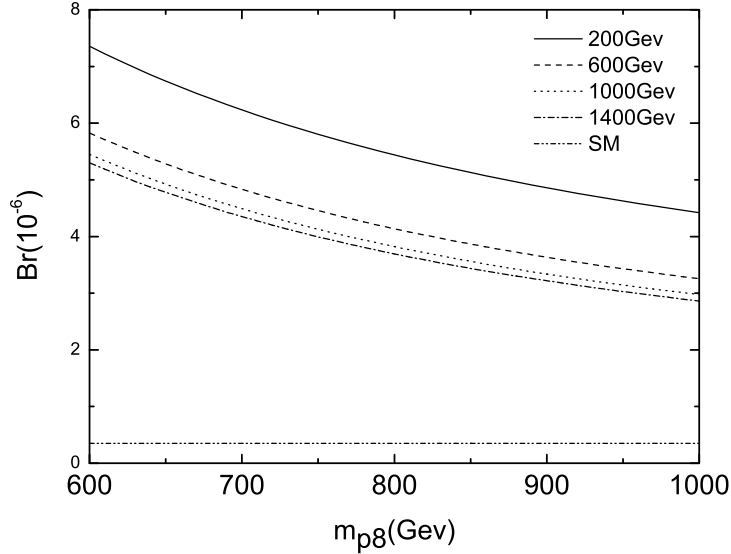}}
\caption
{the $Br(B_{s}\to\gamma\gamma)$ about the mass
of $P_8^\pm$  under different values of $m_{P^\pm}$.} \label{G3.eps}
\end{figure}
\begin{figure}[th]
{\includegraphics{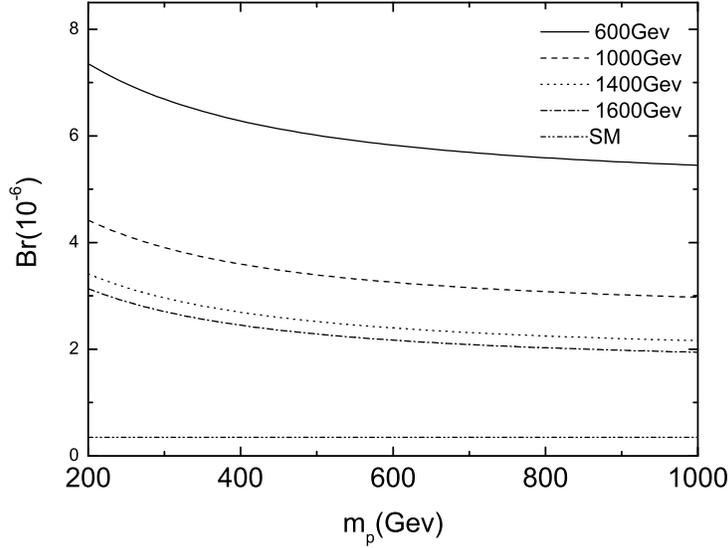}}
\caption
{the $Br(B_{s}\to\gamma\gamma)$ about the mass
of $P^\pm$ under different values of $P_8^\pm$.} \label{g4}
\end{figure}

Fig.3(4) denotes the $Br(B_{s}\to\gamma\gamma)$ about the mass of
$P_8^\pm$ ($P^\pm$) under different values of $m_{P^\pm}$
($P_8^\pm$). From Fig.3 and 4, we find the the curves are much
different from the the SM one. It can be enhanced about 1-2 levels
to the SM prediction in the reasonable region of the masses of PGBs.
This gives the strong new physics signals from the Technicolor
Model. The branching ratio of $B_s\to \gamma\gamma$ decrease along
with the mass of $P_8^\pm$ and $P^\pm$ reduce. This is from the
decoupling theorem that for heavy enough nonstandard boson. When
$m(P^{\pm})$ and $m(P^{\pm}_8)$ have large values, the contributions
from OGTM is small.From the $Eq(16),(17),(18)$ ,we can see the
functions $B$, $D$ and $E$ go to zero, as $x$, $y\to \infty$.The
branching ratio in the Fig.(3) is changed much faster than that in
the Fig.(4).This is because the contribution to $B_s\to\gamma\gamma$
from the color octet $P_8^\pm$ is large when compared with the
contribution from color singlet $P^\pm$.

As a conclusion, the size of contribution to the rare decay of
$B_s\to\gamma\gamma$ from the PGBs strongly depends on the values of
the masses of the charged PGBs. This is quite different from the SM
case. By the comparison of the theoretical prediction with the
current data one can derived out the  the contributions of the PGBs:
$P^\pm$ and $P^\pm_8 $ to $B_s\to\gamma\gamma$ and give the new
physics signals of new physics.

\end{document}